\documentclass[referee,pdflatex,sn-mathphys-num]{sn-jnl}% Math and Physical Sciences Numbered Reference Style
\usepackage{graphicx}%
\usepackage{multirow}%
\usepackage{amsmath,amssymb,amsfonts}%
\usepackage{amsthm}%
\usepackage{mathrsfs}%
\usepackage[title]{appendix}%
\usepackage{xcolor}%
\usepackage{textcomp}%
\usepackage{manyfoot}%
\usepackage{booktabs}%
\usepackage{algorithm}%
\usepackage{algorithmicx}%
\usepackage{algpseudocode}%
\usepackage{listings}%
\usepackage{hyperref}
\usepackage[nameinlink]{cleveref}
\Crefname{equation}{Eq.}{Eqs.}
\Crefname{figure}{Fig.}{Figs.}
\Crefname{section}{Sec.}{Secs.}
\crefrangeformat{figure}{Figs. #3#1#4--#5#2#6}

\usepackage{lmodern}

\usepackage{xr}
\makeatletter

\newcommand*{\addFileDependency}[1]{% argument=file name and extension
\typeout{(#1)}% latexmk will find this if $recorder=0
\@addtofilelist{#1}
\IfFileExists{#1}{}{\typeout{No file #1.}}
}\makeatother

\newcommand*{\myexternaldocument}[1]{%
\externaldocument{#1}%
\addFileDependency{#1.tex}%
\addFileDependency{#1.aux}%
}
\myexternaldocument{SM}

\theoremstyle{thmstyleone}%
\theoremstyle{thmstyletwo}%

\theoremstyle{thmstylethree}%

\newif\ifciteSM
\citeSMfalse

\begin{document}

\title[Article Title]{Single-molecule Scale Nuclear Magnetic Resonance Spectroscopy using a Robust Near-Infrared Spin Sensor }

\author[1]{\fnm{Yu} \sur{Chen}}
%\email{chenyu2018@mail.ustc.edu.cn}

\author*[1,4,5]{\fnm{Qi} \sur{Zhang}}\email{zhq2011@ustc.edu.cn}

\author[1]{\fnm{Yuanhong} \sur{Teng}}
%\email{xassam@mail.ustc.edu.cn}

\author[1]{\fnm{Chihang} \sur{Luo}}
%\email{luoch@mail.ustc.edu.cn}

\author[1]{\fnm{Zhijie} \sur{Li}}
%\email{zjl18@mail.ustc.edu.cn}

\author[1,4]{\fnm{Jinpeng} \sur{Liu}}
%\email{liujinpeng@mail.ustc.edu.cn}

\author[1,3]{\fnm{Ya} \sur{Wang}}
%\email{ywustc@ustc.edu.cn}
\author*[1,2,3,4]{\fnm{Fazhan} \sur{Shi}}\email{fzshi@ustc.edu.cn}

\author[3,6]{\fnm{Jiangfeng} \sur{Du}}

\affil*[1]{\orgdiv{Laboratory of Spin Magnetic Resonance, School of Physical Sciences, Anhui Province Key Laboratory of Scientific Instrument Development and Application}, \orgname{University of Science and Technology of China}, \orgaddress{\city{Hefei}, \postcode{230026}, \country{China}}}

\affil[2]{\orgdiv{Hefei National Research Center for Physical Sciences at the Microscale}, \orgname{University of Science and Technology of China}, \orgaddress{\city{Hefei}, \postcode{230026}, \country{China}}}

\affil[3]{\orgdiv{Hefei National Laboratory}, \orgname{University of Science and Technology of China}, \orgaddress{\city{Hefei}, \postcode{230088}, \country{China}}}

\affil[4]{\orgdiv{School of Biomedical Engineering and Suzhou Institute for Advanced Research}, \orgname{University of Science and Technology of China}, \orgaddress{\city{Suzhou}, \postcode{215123}, \country{China}}}

\affil[5]{\orgdiv{Institute of Quantum Sensing, School of Physics, Institute of Fundamental and Transdisciplinary Research}, \orgname{Zheiiang Key Laboratory of R\&D and Application of Cutting-edge Scientifc Instruments, Zhejiang University}, \orgaddress{\city{Hangzhou}, \postcode{310027}, \country{China}}}

\affil[6]{\orgdiv{State Key Laboratory of Ocean Sensing and School of Physics}, \orgname{Zhejiang University}, \orgaddress{\city{Hangzhou}, \postcode{310058}, \country{China}}}

%%==================================%%
%% sample for unstructured abstract %%
%%==================================%%
\abstract{Nuclear magnetic resonance (NMR) at the single-molecule level with atomic resolution holds transformative potential for structural biology and surface chemistry. Near-surface solid-state spin sensors with optical readout ability offer a promising pathway toward this goal. However, their extreme proximity to target molecules demands exceptional robustness against surface-induced perturbations. Furthermore, life science applications require these sensors to operate in biocompatible spectral ranges that minimize photodamage. In this work, we demonstrate that the PL6 quantum defect in 4H silicon carbide (4H-SiC) can serve as a robust near-infrared spin sensor. This sensor operates at tissue-transparent wavelengths and exhibits exceptional near-surface stability even at depth of 2 nm. Using shallow PL6 centers, we achieve nanoscale NMR detection of proton ($\mathrm{^{1}H}$) spins in immersion oil and fluorine ($\mathrm{^{19}F}$) spins in Fomblin, attaining a detection volume of $\mathrm{(3~nm)^3}$ and a sensitivity reaching the requirement for single-proton spin detection. This work establishes 4H-SiC quantum sensors as a compelling platform for nanoscale magnetic resonance, with promising applications in probing low-dimensional water phases, protein folding dynamics, and molecular interactions.}

\keywords{silicon carbide, color center, nanoscale NMR, quantum sensing}

\maketitle

\section{Introduction}\label{sec:introduction}
Magnetic resonance spectroscopy has profoundly advanced the ability to probe molecular structure and dynamics across chemistry, materials science, and biomedicine \cite{1996-ProteinNMRSpectroscopy-Cavanagh-,2005-BiomedicalEPRPart-Eaton-}. Extending this capability to the single-molecule scale with atomic resolution promises to unlock new frontiers in structural biology and surface chemistry \cite{2024-SinglemoleculeScaleMagnetic-Du-Rev.Mod.Phys.,2024-RoadmapNanoscaleMagnetic-Budakian-Nanotechnology}. Among various solid-state spin sensors, the nitrogen-vacancy (NV) center in diamond has emerged as a leading platform for nanoscale magnetic resonance, enabling detection of single-electron spins and nuclear spin ensembles of single biomolecules under ambient conditions \cite{2015-SingleproteinSpinResonance-Shi-Science,2018-SingleDNAElectronSpin-Shi-NatMethods,2016-NuclearMagneticResonance-Lovchinsky-Science}. Nevertheless, the biology application of NV centers within a few nanometers to surface faces two constraints. First, NV centers require visible-light excitation, which poses risks of photodamage to biological samples \cite{2017-MolecularQuantumSpin-Schlipf-Sci.Adv.,2018-SingleDNAElectronSpin-Shi-NatMethods}. Second, NV centers suffer from charge-state instability within 5 nm to diamond surfaces \cite{2012-SpinPropertiesVery-Ofori-Okai-Phys.Rev.B,2022-DiamondSurfaceEngineering-Janitz-J.Mater.Chem.C}. In this work, we demonstrate that the PL6 center in 4H-SiC serves as an ideal quantum sensor that effectively overcomes the both limitations. 

Using PL6 centers 2–4 nm beneath the surface, we achieve nanoscale nuclear magnetic resonance (NMR) detection, attaining a detection volume of $\mathrm{(3~nm)^3}$ and a sensitivity reaching the requirement for single-proton spin detection. Remarkably, PL6 centers maintain stable operation even at 2 nm below the surface under illumination at 2.2 times the saturation power, demonstrating unprecedented photostability. Furthermore, their near-infrared optical addressability (zero-phonon line at 1038 nm) is compatible with biological tissues, substantially reducing the risk of photodamage to light-sensitive samples compared to visible-wavelength excitation. These advantageous properties—combined with excellent room-temperature spin coherence characteristics \cite{2011-RoomTemperatureCoherent-Koehl-Nature,2022-RoomtemperatureCoherentManipulation-Li-NationalScienceReview,2024-RobustSingleModified-He-NatCommun} and the mature crystal growth and microfabrication technologies available for SiC \cite{2018-ReviewSiCCrystal-Wellmann-Semicond.Sci.Technol.,2023-NovelPhotonicApplications-Ou-Materials}—establish PL6 centers within 5 nm to surface as a highly promising platform for nanoscale magnetic resonance sensing.
\begin{figure*}[h]%
\centering
\includegraphics[]{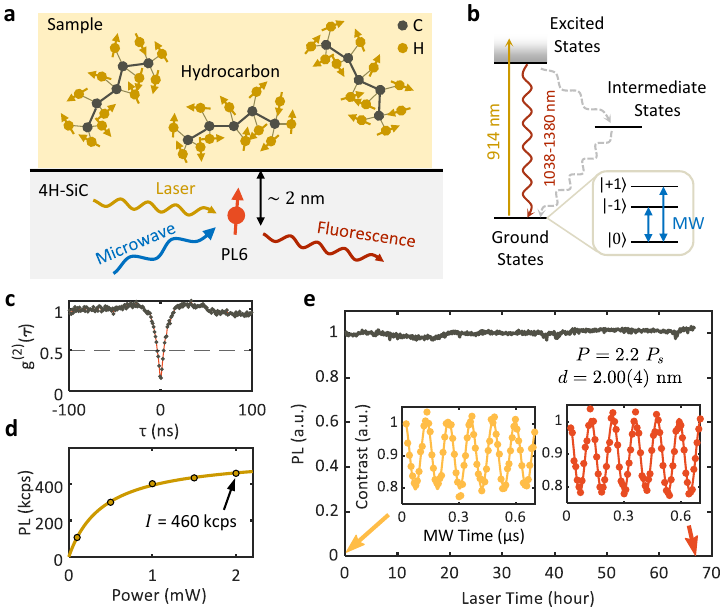}
\caption{\textbf{PL6 color center for nanoscale NMR spectroscopy. } \textbf{a,} Schematic of a shallow PL6 center in 4H-SiC detecting  proton spins within immersion oil placed on the 4H-SiC surface. \textbf{b,} PL6 energy level diagram. \textbf{c-d,} Measured auto-correlation function $g^{(2)}(\tau)$ and saturation curve of a single PL6 emitter. \textbf{e,} Photostability of a shallow PL6 with laser power 2.2 times the saturation power $P_s$. The depth of the PL6 is calibrated with NMR technique \cite{2016-NMRTechniqueDetermining-Pham-Phys.Rev.B}. Insets show Rabi oscillations before (yellow) and after (red) illumination, demonstrating maintained optical spin readout contrast. }\label{fig1}
\end{figure*}
\section{Results}\label{sec:results}
The experimental setup is illustrated in \Cref{fig1}a. The sample was placed on the 4H-SiC surface, and a shallow PL6 color center implanted approximately 2 nm beneath the surface served as the sensor. Experimentally, PL6 features a spin-1 ground-state system that can be coherently manipulated with microwave (\Cref{fig1}b). Optical excitation at 914 nm promotes the system to excited states, followed by radiative decay to ground states with photoluminescence (PL) between 1038‒1380 nm (red arrow in \Cref{fig1}b) or decay via an electron spin–dependent intersystem-crossing pathway through intermediate states (dashed arrows in \Cref{fig1}b) \cite{2015-QuantumEntanglementAmbient-Klimov-Sci.Adv.}. 
Analogous to NV centers in diamond, these relaxation pathways enable optical pumping of the PL6 spins into the $|0\rangle$ state and create spin-dependent PL contrast between the $|\pm1\rangle$ and $|0\rangle$ states, permitting optical readout of the spin state. Under low-flux ion implantation conditions, most PL6 emitters exhibit single-photon emission characteristics, as evidenced by second-order autocorrelation measurements $g^2(0)<0.5$ (\Cref{fig1}c). \Cref{fig1}d illustrates a PL6 center with achieved PL intensity of 460 kcps. The saturation PL intensity for this PL6 is fitted to be 558(31) kcps. Remarkably, a 2-nm-deep PL6 center showed no degradation in fluorescence intensity or optical readout contrast after 60 hours of continuous illumination at 2.2 times the saturation power (\Cref{fig1}e), demonstrating exceptional photostability crucial for practical applications. Moreover, the coherence time of the tested sensor show no degradation after illumination (see Supplementary Materials \ifciteSM \Cref{figS:NMRIllumi}\else Fig. S1\fi).

\begin{figure*}[h]
\centering
\includegraphics{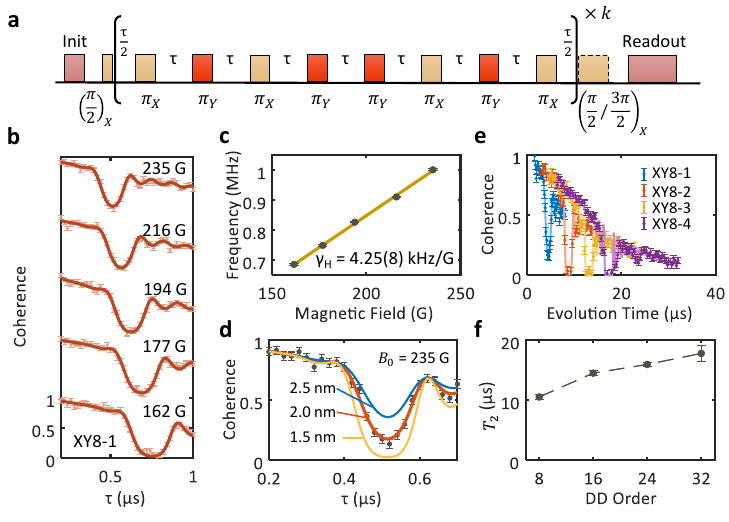}
\caption{\textbf{Nanoscale nuclear magnetic resonance with single shallow PL6 center.} 
\textbf{a,} Schematic illustration of the XY8-$k$ dynamical decoupling pulse sequence employed for NMR detection. 
\textbf{b,} Measured $\mathrm{^1H}$ NMR spectra at various applied magnetic fields, exhibiting characteristic coherence dips. The solid curves represent theoretical fits to the expected NMR lineshape. 
\textbf{c,} Magnetic field dependence of the $\mathrm{^1H}$ resonance frequency, with a linear fit yielding a gyromagnetic ratio $\gamma_\mathrm{H} = 4.25(8)$~kHz/G. 
\textbf{d,} Proton NMR spectrum acquired at $B_0 = 235$ G with the optimal fit (red curve) indicating $d=~$2.00(4) nm depth, compared with theoretically predicted NMR spectra for shallower ($d = 1.5$ nm, golden curve) and deeper ($d = 2.5$ nm, blue curve) sensors.
\textbf{e,}  Proton NMR spectrum measured with varying dynamical decoupling orders at $B_0 = 216$ G.
\textbf{f,} Corresponding coherence times extracted from data in panel (e).
}
\label{fig2}
\end{figure*}

We demonstrate NMR detection of proton spins in immersion oil using an $\text{XY8-}k$ dynamical decoupling (DD) sequence \cite{1990-NewCompensatedCarrPurcell-Gullion-JournalofMagneticResonance1969} (\Cref{fig2}a). The measurement protocol begins by initializing the PL6 spin to the $|0\rangle$ state, followed by a $\pi/2$ pulse that prepares the superposition state $(|0\rangle + |\text{+}1\rangle)/\sqrt{2}$. The subsequent $\text{XY8-}k$ sequence enables AC magnetic field sensing, during which the accumulated phase $\phi$ modifies the quantum state to $(|0\rangle + e^{i\phi}|\text{+}1\rangle)/\sqrt{2}$. Final state readout is achieved through $\pi/2$ or $3\pi/2$ pulse followed by optical detection, providing the coherence $C =\langle e^{i\phi} \rangle\approx e^{-\langle \phi^2 \rangle/2}$.
The $\text{XY8-}k$ sequence serves as a narrow-band filter $G_k(\omega,\tau)$ with transmission local maximum at $\omega=(2m+1)\pi/\tau$ and detection bandwidth $\Delta \omega = 0.222\pi/k\tau$ \cite{2015-NanoscaleNMRSpectroscopy-DeVience-NatureNanotech}. The coherence is given by \cite{2016-NMRTechniqueDetermining-Pham-Phys.Rev.B}:
\begin{equation}
\label{eq:coherence}
  C=e^{-\langle \phi ^2 \rangle/2}
  =\exp \left( -\frac{\gamma^2}{2}  \int_{-\infty}^{+\infty} \frac{d\omega}{2 \pi} G_k(\omega,\tau) S_B(\omega) \right)
\end{equation}
in which $\gamma$ represents the gyromagnetic ratio of PL6 and $S_B(\omega)$ is the power spectrum of the external magnetic field. 

\Cref{fig2}b presents the measured $\mathrm{^1H}$ nuclear magnetic resonance spectra under various  applied magnetic fields $B_0$ aligned with the PL6 axis. Distinct coherence dips are observed at $\tau = \pi/\omega_L$, where $\omega_L$ is the Larmor frequency of $\mathrm{^1H}$ spins. These spectral features originate from statistically polarized proton spins, which produce random transverse magnetization components $\langle M_x\rangle$ and $\langle M_y\rangle$. 
Under the static field $B_0$, these magnetization components precess at the Larmor frequency, generating an oscillating magnetic field $B_{z,\mathrm{H}} = B(\langle M_x\rangle, \langle M_y\rangle)\cos(\omega_L t + \phi)$ at the PL6 sensor location, which results a sharp spectral feature in $S_B(\omega)$ at $\omega_L$. By systematically varying the pulse spacing $\tau$, the transmission maximum of the filter function $G_k(\omega,\tau)$ located at $\pi/\tau$ is tuned. A coherence dip emerges when this maximum aligns with the spectral feature in $S_B(\omega)$ at $\omega_L$, as shown in \Cref{fig2}b.

We extract both the Larmor frequency $\omega_L$ and the RMS magnetic field $B_{\mathrm{RMS,H}}$ of $\mathrm{^1H}$ spins by fitting the NMR spectra in \Cref{fig2}b to a theoretical model (see Supplementary Materials \ifciteSM\Cref{secS:Dynamical Decoupling Sequence}\else Sec. S1\fi). \Cref{fig2}c displays the linear relationship $\omega_L/2\pi=\gamma_\mathrm{H} B_0$, where the fitted slope $\gamma_\mathrm{H} = 4.25(8)$~kHz/G matches the known proton gyromagnetic ratio 4.26 kHz/G. The measured $B_{\mathrm{RMS,H}}$ of $3.7(1)~\mathrm{\mu T}$, combined with the known proton density in immersion oil (69.5~$\mathrm{nm^{-3}}$) \cite{2023-SubnanoteslaSensitivityNanoscale-Zhao-NationalScienceReview}, yields a calculated sensor depth of 2.00(4)~nm (see Supplementary Materials \ifciteSM\Cref{secS:Depth Calibration}\else Sec. S3\fi). \Cref{fig2}d compares the experimental NMR spectrum at $B_0=235~\mathrm{G}$ with simulations for different sensor depths (1.5 and 2.5 nm), confirming the optimal fit.

\begin{figure}[h]
\centering
\includegraphics[]{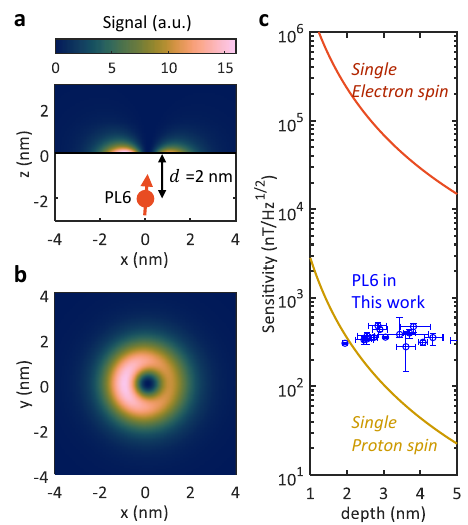}
\caption{\textbf{Detection volume and sensitivity characterization.} \textbf{a-b,} Simulated magnetic field contributions showing the detection volume of a 2~nm-deep PL6 sensor. \textbf{c,} Measured sensitivity of fourteen shallow PL6 centers (blue dots) compared with theoretical thresholds for single-spin detection, including the requirements for ENDOR detection of single proton spin (yellow line) and DEER detection of single electron spin (red line) at SNR = 1 with 1 second integration time.}
\label{fig3}
\end{figure}
To characterize the spacial resolution of our detection, we perform numerical simulations of the NMR signal distribution with sensor positioned at (0, 0, -2) nm. \Cref{fig3}a and \Cref{fig3}b present the signal simulated distribution in the XZ and XY planes. At this shallow depth of 2~nm, the PL6 sensor predominantly detects signals from approximately 2500 proton spins, which has a statistical polarization equivalent to $\sim$50 polarized proton nuclei, accounting for $\sim$70\% of the total detected signal. This corresponds to an effective detection volume of $\mathrm{(3~nm)^3}$, corresponding to the single-molecule scale.

We further characterized the detection sensitivity by analyzing fourteen shallow PL6 centers. The sensitivities were calculated using the coherence time measured with XY8-2 sequence and the measured average numbers of photons per single readout (see Supplementary Materials \ifciteSM\Cref{secS:Sensitivity,secS:Single Spin Detection} \else Secs. S5--6 \fi for details). As shown in \Cref{fig3}c, the measured sensitivities are presented alongside the threshold values needed to achieve a signal-to-noise ratio (SNR) of 1 in 1 second integration time for both double electron-electron resonance (DEER) \cite{2015-SingleproteinSpinResonance-Shi-Science} detection of single electron spin and electron-nuclear double resonance (ENDOR) \cite{2013-NanoscaleNuclearMagnetic-Mamin-Science} detection of single proton spin. 
The PL6 sensors exhibit a sensitivity of $\sim\mathrm{350~nT/Hz^{1/2}}$ at depths of 2–4 nm, sufficient for single-electron spin detection, with one case ($\mathrm{307(9)~nT/Hz^{1/2}}$ at depth 2.0 nm) reaching the requirement for single-proton detection (\Cref{fig3}c).

\begin{figure}[h]
\centering
\includegraphics[width=\linewidth]{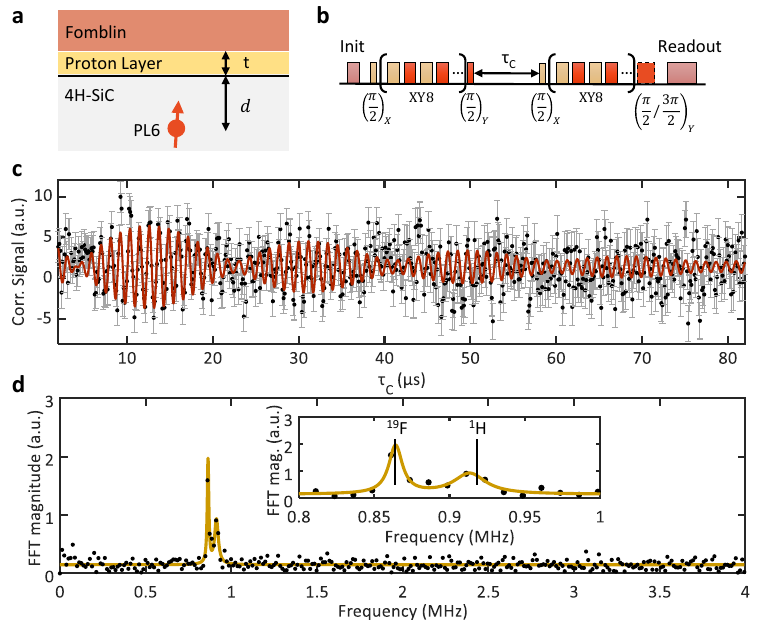}
\caption{\textbf{Multi-species nanoscale NMR spectroscopy.} \textbf{a,} Experimental schematic for detecting $\mathrm{^{19}F}$ in Fomblin and surface-adsorbed $\mathrm{^{1}H}$ on 4H-SiC surface. \textbf{b,} XY8 correlation spectroscopy pulse sequence. \textbf{c,} Correlation signal showing oscillations from both $\mathrm{^{1}H}$ (proton layer) and $\mathrm{^{19}F}$ (Fomblin) at $B_0 = 215.8$~G. The solid line represents a fit using double decaying cosine model, yielding $T^{(^1\mathrm{H})} _{\mathrm{coor}}=27(5) ~\mathrm{\mu s}$ and $T^{(^{19}\mathrm{F})} _{\mathrm{coor}}=55(10) ~\mathrm{\mu s}$.. \textbf{d,} Fourier transform of the correlation signal revealing distinct $\mathrm{^{1}H}$ and $\mathrm{^{19}F}$ peaks. Inset displays a magnified view with two-Lorentzian fitting (solid curve), yielding measured FWHM linewidths of $26(5)$~kHz for $\mathrm{^{1}H}$ and $10(4)$~kHz for $\mathrm{^{19}F}$. Vertical lines indicate the expected Larmor frequencies for both nuclear species at this magnetic field strength. }
\label{fig4}
\end{figure}
Finally, we conducted multi-species nanoscale NMR measurements on a Fomblin (Fomblin Y HVAC 140/13) sample deposited on a 4H-SiC surface (Fig.~\ref{fig4}a). The experiments utilized a correlation spectroscopy sequence (Fig.~\ref{fig4}b) that provides two significant advantages compared to conventional dynamical decoupling sequences: (1) elimination of harmonic artifacts in the signal and (2) enhanced spectral resolution ($\sim 1/T_1$) that surpasses the decoherence-limited resolution ($\sim 1/T_2$) of standard dynamical decoupling methods \cite{2013-HighresolutionCorrelationSpectroscopy-Laraoui-NatCommun,2015-SpuriousHarmonicResponse-Loretz-Phys.Rev.X,2016-OneTwoDimensionalNuclear-Boss-Phys.Rev.Lett.}.
The correlation spectroscopy signal obtained from the Fomblin/4H-SiC system exhibits two distinct oscillation frequencies (Fig.~\ref{fig4}c). Fourier transform analysis (Fig.~\ref{fig4}d) confirms that these frequencies correspond to the Larmor frequencies of $\mathrm{^{19}F}$ and $\mathrm{^{1}H}$ at the applied magnetic field of 215.8~G (inset of Fig.~\ref{fig4}d). Since Fomblin contains no hydrogen in its molecular structure, we attribute the proton signal to a surface-adsorbed hydrocarbon or water layer on the 4H-SiC substrate. Additional characterization using dynamical decoupling sequences reveals a PL6 sensor depth of $d = 3.6(2)$~nm and a measured thickness of $t = 0.8(3)$~nm for the proton layer in 4H-SiC surface (see Supplementary Materials \ifciteSM\Cref{figS:FomblinDD780G} \else Fig. S2 \fi for experimental details). Such surface proton layers have also been observed on diamond surfaces with typical thicknesses of $0.8(1)$~nm \cite{2015-NanoscaleNMRSpectroscopy-DeVience-NatureNanotech}, likely originating from adsorbed water molecules \cite{2024-DiscoveryAnomalousNonevaporating-Li-}.

\section{Discussion}\label{sec:discussion}
In summary, we have demonstrated multi-species nanoscale NMR spectroscopy of $\mathrm{^{1}H}$ and $\mathrm{^{19}F}$ nuclei using shallow PL6 centers in 4H-SiC. The shallow PL6 sensors show great robustness under illumination, and the achieved sensitivity of  $\mathrm{307~nT/Hz^{1/2}}$ at depths of 2 nm highlights the exceptional potential of PL6 centers for single spin spectroscopy. This sensitivity could be further improved by implementing advanced techniques such as repetitive readout with nuclear spin ancillae, spin-to-charge conversion readout schemes, and collection efficiency enhancement via nano-structures \cite{2020-SensitivityOptimizationNVdiamond-Barry-Rev.Mod.Phys.,2016-NuclearMagneticResonance-Lovchinsky-Science,2015-EfficientReadoutSingle-Shields-Phys.Rev.Lett.,2021-HighfidelitySingleshotReadout-Zhang-NatCommun,2017-ScalableQuantumPhotonics-Radulaski-NanoLett.}. In Supplementary Materials, we also demonstrate another point defect, PL5 in 4H-SiC, can serve as a robust nanoscale NMR sensor at a depth of 1.8(1) nm. Given the outstanding photostability and sensitivity of these shallow centers, we anticipate that sub-nanometer centers with preserved properties in 4H-SiC could be realized. Such optimized centers would provide stronger signal amplitudes, achieving magnetic resonance imaging of single molecule at atomic resolution.

\section{Methods}\label{sec:methods}

\textbf{Sample preparation.} The samples were fabricated from wafers comprising a 12.5-$\mu$m-thick intrinsic epitaxial layer of single-crystal 4H-SiC grown on a $4^\circ$ off-axis N-type 4H-SiC substrate. For the measurements presented in \Cref{fig1}c--d, we used a sample implanted with 60-keV $^{14}\text{N}^{+}$ ions at a dose of $10^{10}~\text{cm}^{-2}$, followed by thermal annealing at 1000$^\circ$C for 30 minutes in vacuum. \Cref{fig1}e and \crefrange{fig2}{fig4} experimental data were obtained from a separate sample that was implanted with 3-keV $^{14}\text{N}^{+}$ ions at the same dose and subsequently annealed at 1050$^\circ$C for 30 minutes in vacuum. Fomblin (Fomblin Y HVAC 140/13) or immersion oil (IMMOILF30CC) of the objective is placed upon the 4H-SiC surface.

\textbf{Optical measurements.} The measurements were conducted on a home-built scanning confocal microscope with an infrared oil objective with an NA of 1.45 (Nikon, CFI Plan Apochromat Lambda D). A 914-nm CW laser, filtered by a shortpass filter (Thorlabs, FESH950), was used to excite those color centers. A dichroic beamsplitter (Semrock, Di02-R98025×36) was then used to separate the laser and fluorescence signals. The fluorescence signals filtered by a 980-nm longpass filter (Semrock, BLP01-980R-25) were coupled to a single-mode fibre and then guided to a superconducting nanowire single-photon detector (SNSPD).

\backmatter

\bmhead{Supplementary information}
Supplementary information is available in the attachment of the paper.

\bmhead{Acknowledgments}
This work was supported by the National Natural Science Foundation of China (Grant Nos. T2125011, 12174377), the CAS (Grant Nos. YSBR-068),  Innovation Program for Quantum Science and Technology (Grant Nos. 2021ZD0302200, 2021ZD0303204, 2023ZD0300100), New Cornerstone Science Foundation through the XPLORER PRIZE, Science and Technology Department of Zhejiang Province (2025C01041) and the Fundamental Research Funds for the Central Universities (226-2024-00142). This work was partially carried out at the USTC Center for Micro and Nanoscale Research and Fabrication.

\bibliography{main_bibliography}

\end{document}

%TODO: Modify the title of SM
\title[Article Title]{Supplemental Material to ``Single-molecule Scale Nuclear Magnetic Resonance Spectroscopy using a Robust Near-Infrared Spin Sensor"}
\maketitle
\section{Dynamical Decoupling Sequence\label{secS:Dynamical Decoupling Sequence}}
During the dynamical decoupling measurement sequence, the NV spin coherence accumulates some phase $\phi$, the remaining coherence is given by \cite{2016-NMRTechniqueDetermining-Pham-Phys.Rev.B}
\begin{equation}
    C(t) = e^{-\langle \phi ^2 \rangle/2}
  =\exp \left( -\frac{\gamma^2}{2}  \int_{-\infty}^{+\infty} \frac{d\omega}{2 \pi} 
  G(\omega,\tau) S_B(\omega) \right)
  \label{eqS:ct}
\end{equation}
in which $\gamma$ is the gyro-magnetic ratio of PL6, $S_B(\omega)$ is the power spectrum of the outer magnetic field. Separating the background spin decoherence out, \Cref{eqS:ct} becomes
\begin{equation}
     C(t) = \exp \left( -\frac{\gamma^2}{2}  \int_{-\infty}^{+\infty} \frac{d\omega}{2 \pi} 
  G(\omega,\tau) S_{B,\mathrm{N}}(\omega) \right)
  \exp \left(
  -\left( \frac{t}{T_2}\right)^n
  \right)
  \label{eqS:ct1}
\end{equation}
$S_{B,N}(\omega)$ is the power spectrum of the outer magnetic field caused by sample nuclear spins. The band filter $G(\omega,\tau)$ for XY8-k sequence is \cite{2016-NMRTechniqueDetermining-Pham-Phys.Rev.B}
\begin{equation}
    G(\omega,\tau) = 
\frac{16\sin^{4}\left(\frac{w\tau }{4}\right)\sin^{2}\left(\frac{Nw\tau }{2}\right)}{\omega ^{2}\cos^{2}\left(\frac{w\tau }{2}\right)}
\label{eqS:Sg}
\end{equation}
where $N$ is the number of $\pi$-pulses and $\tau$ is the free precession time between $\pi$-pulses. 

For NMR signal of $\mathrm{^{1}H}$ in  immersion oil measured by XY8 sequence,  the magnetic field power spectrum of $\mathrm{^{1}H}$ spins is \cite{2016-NMRTechniqueDetermining-Pham-Phys.Rev.B}
\begin{equation}
\begin{split}
         S_{B,\mathrm{N}}(\omega) = &\pi B_{\mathrm{RMS,H}}^2 (\delta(\omega-\omega_{L,\mathrm{H}})+\delta(\omega+\omega_{L,\mathrm{H}})) \\
\end{split}
\label{eqS:SBNH}
\end{equation}
where $\omega_{L,\mathrm{H}} $ is the Larmor frequency of $\mathrm{^{1}H}$. The parameters  $\omega_{L,\mathrm{H}} $, $B_{\mathrm{RMS,H}}$ were determined by fitting the NMR signal with \Cref{eqS:ct1,eqS:Sg,eqS:SBNH}.

For NMR signal of $\mathrm{^{1}H}$ in proton layer and $\mathrm{^{19}F}$ in Fomblin measured by XY8-k sequence, 
\begin{equation}
\begin{split}
         S_{B,\mathrm{N}}(\omega) = &\pi B_{\mathrm{RMS,H}}^2 (\delta(\omega-\omega_{L,\mathrm{H}})+\delta(\omega+\omega_{L,\mathrm{H}})) \\
    & +     \pi B_{\mathrm{RMS,F}}^2 (\delta(\omega-\omega_{L,\mathrm{F}})+\delta(\omega+\omega_{L,\mathrm{F}})) 
\end{split}
\label{eqS:SBNHF}
\end{equation}
where $\omega_{L,\mathrm{H}} $ is the Larmor frequency of $\mathrm{^{1}H}$ , $\omega_{L,\mathrm{F}}$ is the Larmor frequency of $\mathrm{^{19}F}$. The parameters $\omega_{L,\mathrm{H}}  $, $\omega_{L,\mathrm{F}}  $, $B_{\mathrm{RMS,H}}$, $B_{\mathrm{RMS,F}}  $ were determined by fitting the NMR signal with \Cref{eqS:ct1,eqS:Sg,eqS:SBNHF} .

\section{Stable NMR performance of PL6 sensor under Illumination\label{secS:Stable NMR performance of PL6 sensor under Illumination}}
\begin{figure*}
    \centering
    \includegraphics[]{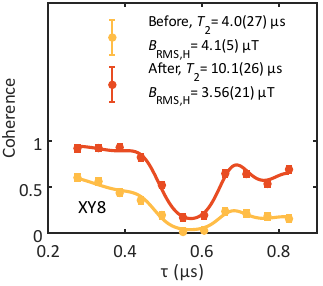}
    \caption{\textbf{NMR signal before(yellow) and after(red) illumination.} The signal is fitted with \Cref{eqS:ct1,eqS:Sg,eqS:SBNH}. Obtained $T_2, B_{\mathrm{RMS,H}}$ are displayed.}
    \label{figS:NMRIllumi}
\end{figure*}

We measured NMR signal of $\mathrm{^{1}H}$ in  immersion oil with 2 nm-deep PL6 sensor before and after illumination with XY8 sequence (\Cref{figS:NMRIllumi}), the coherence raised after illumination in the tested PL6 sensor and the obtained signal from  $\mathrm{^{1}H}$  remained unchanged.

\section{Depth Calibration\label{secS:Depth Calibration}}
% TODO: add the SRIM Result here
 The $B_{\mathrm{RMS}}$ created by a layer of spin-1/2 nuclei with density $\rho$ and thickness $t$ is \cite{2016-NMRTechniqueDetermining-Pham-Phys.Rev.B}
\begin{equation}
B_{\mathrm{RMS}}^2  
 =\rho\left(\frac{\mu_0 \hbar \gamma_n}{4 \pi}\right)^2\left(\frac{\pi\left[8-3 \sin ^4(\alpha)\right]}{128 }\right) 
\left(  \frac{1}{d^3} - \frac{1}{(d+t)^3}  \right) 
\label{eqS:Brms}
\end{equation}
in which $\gamma_n$ is the gyromagnetic ratio of the nuclei, $\alpha =4^\circ$ is the angle between the PL6 axis and the vector normal to the diamond surface, $d$ is the distance between PL6 and semi-infinite layer. 

For NMR measurement of $\mathrm{^{1}H}$ in immersion oil,  we assume the $\mathrm{^{1}H}$ density in proton layer is approximately the same as in immersion oil so the immersion oil and the proton layer can be viewed as one layer. The thickness of  immersion oil is infinity ($t\xrightarrow{}+\infty$) and the $\mathrm{^{1}H}$ density in it is assumed to be $\rho_\mathrm{H} = 69.5 ~\mathrm{nm^{-3}}$ \cite{2023-SubnanoteslaSensitivityNanoscale-Zhao-NationalScienceReview}, we have $B_{\mathrm{RMS,H}}^2$ in \Cref{eqS:BrmsH} as
\begin{equation}
B_{\mathrm{RMS,H}}^2  
 =\rho_\mathrm{H}\left(\frac{\mu_0 \hbar \gamma_\mathrm{H}}{4 \pi}\right)^2\left(\frac{\pi\left[8-3 \sin ^4(\alpha)\right]}{128 }\right) 
 \frac{1}{d^3} 
\label{eqS:BrmsH}
\end{equation}
The depth calibration is carried by fitting the XY8 signal of $\mathrm{^{1}H}$ in  immersion oil. With $B_{\mathrm{RMS,H}}$ obtained from fitting, the distance $d$ between PL6 and $\mathrm{^{1}H}$ layer can be calculated as
\begin{equation}
  d  = \left( 
  \rho _\mathrm{H}
 \left(\frac{\mu_0 \hbar \gamma_{\mathrm{H}}}{4 \pi}\right)^2
 \left(\frac{\pi\left[8-3 \sin ^4(\alpha)\right]}{128 }\right) 
\frac{1}{B_{\mathrm{RMS,H}}^2  } 
\right)^\frac{1}{3}
\end{equation}
We attribute it as the depth of PL6.

%With $B_{\mathrm{RMS}}$, later using $\mathrm{^{1}H}$ density $\rho = 69.5 ~\mathrm{nm^{-3}}$ and \Cref{eqS:Brms} we can get the depth of PL6. It's worth mentioning in this case we simply assume the $\mathrm{^{1}H}$ density in proton layer is approximately the same as in immersion oil.
\section{Proton Layer Thickness\label{secS:Proton Layer Thickness}}
For  NMR signal of $\mathrm{^{1}H}$ in proton layer and $\mathrm{^{19}F}$ in Fomblin measured by XY8 sequence, we have $ B_{\mathrm{RMS,H}}^2  ,B_{\mathrm{RMS,F}}^2  $ in \Cref{eqS:BrmsHF} as
\begin{equation}
\begin{split}
    B_{\mathrm{RMS,H}}^2  
 &=\rho _\mathrm{H}
 \left(\frac{\mu_0 \hbar \gamma_{\mathrm{H}}}{4 \pi}\right)^2
 \left(\frac{\pi\left[8-3 \sin ^4(\alpha)\right]}{128 }\right) 
\left(  \frac{1}{d^3} - \frac{1}{(d+t)^3}  \right) \\
B_{\mathrm{RMS,F}}^2  
& =\rho_\mathrm{F}
\left(\frac{\mu_0 \hbar \gamma_{\mathrm{F}}}{4 \pi}\right)^2\left(\frac{\pi\left[8-3 \sin ^4(\alpha)\right]}{128 }\right) 
\frac{1}{(d+t)^3}  
\end{split}
\label{eqS:BrmsHF}
\end{equation}
where $d$ is the depth of PL6 and $t$ is the thickness of proton layer. With $B_{\mathrm{RMS,H}},B_{\mathrm{RMS,F}}$ obtained from fitting, the depth $d$ and thickness $t $ can be calculated as
\begin{equation}
\begin{split}
      d  &= \left( 
  \rho _\mathrm{H}
 \left(\frac{\mu_0 \hbar \gamma_{\mathrm{H}}}{4 \pi}\right)^2
 \left(\frac{\pi\left[8-3 \sin ^4(\alpha)\right]}{128 }\right) 
\frac{1}{
B_{\mathrm{RMS,H}}^2  
+B_{\mathrm{RMS,F}}^2  
(\frac{\gamma_{\mathrm{H}}}{\gamma_{\mathrm{F}}})^2
\frac{\rho_{\mathrm{H}}}{\rho_{\mathrm{F}}}
}
\right)^\frac{1}{3}
\\
t  & =\left( 
\rho_\mathrm{F}
\left(\frac{\mu_0 \hbar \gamma_{\mathrm{F}}}{4 \pi}\right)^2\left(\frac{\pi\left[8-3 \sin ^4(\alpha)\right]}{128 }\right) 
\frac{1}{B_{\mathrm{RMS,F}}^2}  
\right)^\frac{1}{3} - d
\end{split}
\label{eqS:Cal_d_t}
\end{equation}
The $\mathrm{^{19}F}$ density in Fomblin (Fomblin Y HVAC 140/13) is estimated as $\rho_\mathrm{F}= 42 ~ \mathrm{nm^{-3}}$. The $\mathrm{^{1}H}$ density in proton layer is assumed to be $\rho_\mathrm{H} = 69.5 ~\mathrm{nm^{-3}}$ \cite{2023-SubnanoteslaSensitivityNanoscale-Zhao-NationalScienceReview}. 
% TODO: add the calculation here to the thickness calculation
\begin{figure*}[h]
    \centering
    \includegraphics{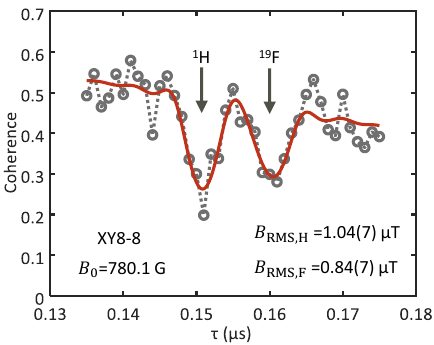}
    \caption{\textbf{NMR signal of $\mathrm{^{1}H}$ in proton layer and $\mathrm{^{19}F}$ in Fomblin measured by XY8-8 sequence.} Fitting the NMR signal with \Cref{eqS:ct1,eqS:Sg,eqS:SBNHF} gives $B_{\mathrm{RMS,H}}=1.04\pm0.07 ~\mathrm{\mu T}$ and $B_{\mathrm{RMS,F}} =0.84\pm0.07 ~\mathrm{\mu T} $. Further using \Cref{eqS:Cal_d_t}, the sensor depth is given by $d=3.6\pm0.2 ~\mathrm{nm}$ and the thickness of proton layer is given by $t=0.8\pm0.3 ~\mathrm{nm}$. }
    \label{figS:FomblinDD780G}
\end{figure*}

%using $\mathrm{^{1}H}$ density $\rho_\mathrm{H} = 69.5 ~\mathrm{nm^{-3}}$ and $\rho_\mathrm{F} =30 ~\mathrm{nm^{-3}}$.   

\section{Sensitivity\label{secS:Sensitivity}}
In optical readout of the population of $|0\rangle$ population $P$, the uncertainty is given by
\begin{equation}
\sigma _{P} \approx \frac{1}{\sqrt{T/( \tau +t_{IR} )}}\frac{1}{C\sqrt{\alpha _{\text{avg}}}}
\end{equation}
where $T$ is the total experiment time, $\tau$ is the integral time, $t_{IR}=t_{I} +t_{R}$ is the sum of the time for initialization and readout, $ C=\frac{\alpha _{0} -\alpha _{1}}{\alpha _{\text{avg}}}$ is the optical readout contrast and $\alpha _{\text{avg}}=(\alpha _{0} +\alpha _{1})/2$ is the average collection photon for single readout. 

\begin{figure*}
    \centering
    \includegraphics[]{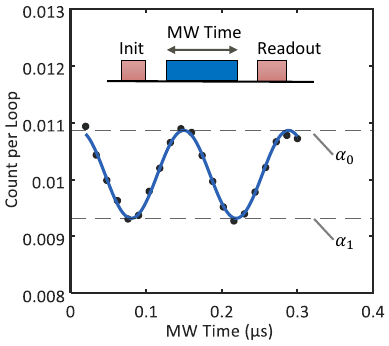}
    \caption{\textbf{Optically detected Rabi oscillations.} Measured Rabi oscillation curve of the PL6 center. The data are fitted with $(\alpha_0+\alpha_1)/2+(\alpha_0-\alpha_1)\cos(\omega t+\phi)/2$ function (solid line) to extract the average single readout collection photon $\alpha_0$, $\alpha_1$ for $|0\rangle$ and $|\pm1\rangle$ state.}
    \label{figS:PhotonCountRabi}
\end{figure*}

For polarization signal, we readout out the imaginary part of the coherence, the population is given out by 
\begin{equation}
P=\frac{1+C_{\tau }\sin \varphi }{2} \approx \frac{1}{2} +\frac{1}{2} C_{\tau }( \gamma B\tau )
\label{eqS:P}
\end{equation}
where $\displaystyle C_{\tau } =\exp\left( -( \tau /T_{2})^{n}\right)$ is coherence remain with background spin decoherence. 
Thus the uncertainty for $B$ we can reach in total experiment time $T$ is
\begin{gather}
 \sigma _{B} =\frac{2\sigma _{P}}{C_{T}( \gamma _{e} \tau )} 
\end{gather}
The sensitivity for $B$ is given by
\begin{equation}
\eta _{B} =\sigma _{B}\sqrt{T} =\frac{2}{\gamma \exp\left( -( \tau /T_{2})^{n}\right)}\frac{1}{\tau ^{1/2}}\sqrt{\frac{\tau +t_{I}{}_{R}}{\tau }}\frac{1}{C\sqrt{\alpha _{\text{avg}}}}
\label{eqS:etaB}
\end{equation}
For stretch factor $n=1$, the best sensitivity is obtained with 
\begin{equation}
\tau _{m} =\frac{1}{4}\left(\sqrt{T_{d}^{2} +12T_{2} t_{IR} +4t{_{IR}}^{2}} +T_{2} -2t_{IR}\right)
\label{eqS:taum_sin}
\end{equation}
For each PL6, the optical readout contrast $C$ and average collection photon $\alpha _{\text{avg}}$ are extracted from optically detected Rabi oscillations (\Cref{figS:PhotonCountRabi}), while  $T_{2}$ is characterized by XY8-2 dynamical decoupling sequences. Initialization and readout time in experiment is $t_{IR}=1 ~\mathrm{\mu s}$. The stretch factor is assumed to be $n=1$. With all the measurement, the sensitivity for each PL6 is calculated with \Cref{eqS:etaB,eqS:taum_sin}.

\section{Single Spin Detection\label{secS:Single Spin Detection}}
For single $I=1/2$ spin placed at $-r\mathbf{n}$ (r is the distance between the sensor and the spin, $\mathbf{n}$ is the unit vector pointing from the sensor to the spin). We choose the orientation of PL6 as the z axis, then the z component of its magnetic field at PL6 sensor is 
\begin{equation}
B_{z} =\frac{\mu _{0} \gamma_s \hbar }{4\pi }\frac{3(\boldsymbol{I} \cdotp \mathbf{n}) n_{z} -I_{z}}{r^{3}}
\end{equation}
where $\gamma_s$ is the gyromagnetic ratio of the spin and $\mathbf{I}$ is the spin operator. Assuming the spin weak coupling the sensor, the spin evolution is dominated  by $H_s=\gamma_sB_0I_z =\omega I_z$, and the correlation of $B_z$ component is 
\begin{equation}
\begin{aligned}
\langle B_{z} (t)B_{z} (0)\rangle  & =\mathrm{Tr} (B_{z} (t)B_{z} (0)\rho )\\
 & =\mathrm{Tr}\left( e^{i\omega tI_{z}} B_{z} (0)e^{-i\omega tI_{z}} B_{z} (0)\rho \right)\\
 & =\frac{b^{2}}{4}\left[\left( 1-3n_{z}^{2}\right)^{2} +9\left( 1-n_{z}^{2}\right) n_{z}^{2}\cos \omega t\right]
\end{aligned}
\end{equation}
with $\displaystyle b=\frac{\mu _{0} \gamma _{s} \hbar }{4\pi r^{3}}$.  For DEER detection of single electron spin or ENDOR detection of a proton spin, the effective $B$ in \Cref{eqS:P} is 
\begin{equation}
B =\sqrt{\frac{b^{2}}{4}\left( 1-3n_{z}^{2}\right)^{2}} 
\end{equation}
For PL6 with depth $d$ and small angle $\alpha =4^\circ$, the signal of a spin in 4H-SiC surface reaches maximum with $n_z\approx1$, this gives out the maximum signal strength $B$ of a single spin
\begin{equation}
B_{\mathrm{max}}\approx \left(\frac{\mu _{0} \gamma _{s} \hbar }{4\pi d^{3}}\right)
\end{equation}
To obtain signal-to-noise ratio SNR=1 in 1 s, the needed sensitivity is 
\begin{equation}
\eta =B_{\mathrm{max}}\sqrt{1\ \mathrm{s}} =
\left(\frac{\mu _{0} \gamma _{s} \hbar }{4\pi d^{3}}\right) /\mathrm{Hz}^{1/2}
\end{equation}

\section{Correlation sequence}
Here the pulse sequence is given by $\displaystyle \left(\frac{\pi }{2}\right)_{X} -(\text{XY8-k}) -\left(\frac{\pi }{2}\right)_{Y} -\tau _{C} -\left(\frac{\pi }{2}\right)_{X} -(\text{XY8-k}) -\left(\frac{\pi }{2}\right)_{Y}$. Upon evolution, where $\tau$ is  the duration of the XY8-k interpulse interval and $\tau _{C}$ is the time separation between two XY8-k. Treat PL6 as a effective two-level system and ignore the relaxation of the sensor ($T_1\xrightarrow{}+\infty$), the density operator after correlation sequence takes the form \cite{2013-HighresolutionCorrelationSpectroscopy-Laraoui-NatCommun}
\begin{equation}
\begin{aligned}
\rho (2N\tau +\tau _{C} )=\frac{1}{2}\{1+\sigma _{\mathrm{y}}\cos \phi _{1}\cos \varphi _{3} -\sigma _{\mathrm{x}} & (\sin \phi _{1}\cos \phi _{2} +\sin \varphi _{3}\cos \phi _{1}\sin \phi _{2}) +\\
+ & \sigma _{\mathrm{z}}(\sin \phi _{1}\sin \phi _{2} -\sin \varphi _{3}\cos \phi _{1}\cos \phi _{2})\} .
\end{aligned}
\end{equation}
where $N$ is the number of $\pi$-pulses in XY8-k sequence and
\begin{gather}
\begin{aligned}
\phi _{1} & =\int _{0}^{N\tau } g_{N}( t) B( t) \ \mathrm{d} t
\end{aligned} ,\ \begin{aligned}
\phi _{2} & =\int _{0}^{N\tau } g_{N}( t) B( t+N\tau +\tau _{C}) \ \mathrm{d} t
\end{aligned} ,\ \\
\begin{aligned}
\varphi _{3} & =\int _{N\tau }^{N\tau +\tau _{C}} B( t) \ \mathrm{d} t
\end{aligned} \  \notag
\end{gather}
\begin{equation}
g_N( t) =\begin{cases}
\gamma( -1)^{m} & ,t\in \left[ m\tau ,m\tau +\frac{\tau }{2}\right]\\
\gamma( -1)^{m+1} & ,t\in \left[ m\tau +\frac{\tau }{2} ,( m+1) \tau \right]
\end{cases} ,m=0,1,\dotsc N-1
\end{equation}
With weak signal approximation $\phi_1,\phi_2\ll1$, the final polarization at z axis is
\begin{equation}
\begin{aligned}
S(t) & =\left< \operatorname{Tr}\{\rho (2N\tau +\tau _{C} )\sigma _{z}\}\right> \\
 & =\langle \sin \phi _{1}\sin \phi _{2} -\sin \varphi _{3}\cos \phi _{1}\cos \phi _{2} \rangle \\
% & =\langle \sin \phi _{1}\sin \phi _{2} \rangle \\
% & =\frac{1}{2} \langle \cos( \phi _{1} -\phi _{2}) -\cos( \phi _{1} +\phi _{2}) \rangle \\
% & \approx \frac{1}{2}\left( e^{-\langle ( \phi _{1} -\phi _{2})^{2} \rangle /2} -e^{-\langle ( \phi _{1} +\phi _{2})^{2} \rangle /2}\right)\\
% & =e^{-\left( \langle \phi _{1}^{2} \rangle +\langle \phi _{2}^{2} \rangle \right) /2}\sinh \langle \phi _{1} \phi _{2} \rangle \\
% & =e^{-\langle \phi ^{2} \rangle }\sinh \langle \phi _{1} \phi _{2} \rangle 
& \approx \langle \phi _{1} \phi _{2} \rangle 
\end{aligned}
\label{eqS:st}
\end{equation}
%$\langle \phi ^{2} \rangle$ is the accumulated phase variance in XY8-k sequence. The first term is equal to the square of the coherence after XY8-k sequence and remain a constant in our correlation measurement. Usually \Cref{eqS:st} will be further simplified with weak signal approximation $\phi_1,\phi_2\ll1$ and becomes $S(t) \approx e^{-\langle \phi ^{2} \rangle } \langle \phi _{1} \phi _{2}\rangle $

 For NMR signal of $\mathrm{^{1}H}$ in proton layer and $\mathrm{^{19}F}$ in Fomblin measured by correlation sequence, $\langle \phi _{1} \phi _{2} \rangle $ should contains two term oscillating with the Larmor frequency of $\mathrm{^{1}H}$ in proton layer and $\mathrm{^{19}F}$ in Fomblin. This can be modeled as 
\begin{equation}
\begin{split}
%       S(t)& = e^{-\langle \phi ^{2} \rangle } \langle \phi _{1} \phi _{2} \rangle\\
%  &=e^{-\tau _{C} \Gamma _{\mathrm{H}}} 
 S(t)& =  \langle \phi _{1} \phi _{2} \rangle\\
  &=e^{-\tau _{C}/T^{(^{1}\mathrm{H})} _{\mathrm{coor}}} 
  A _{\mathrm{H}} 
  \ \cos( \omega _{L\mathrm{,H}} \tau _{C} +\phi_\mathrm{H}) 
  + e^{-\tau _{C} /  T^{(^{19}\mathrm{F})} _{\mathrm{coor}}}
  A _{\mathrm{F}} 
  \ \cos( \omega _{L,\mathrm{F}} \tau _{C} +\phi_\mathrm{F})
\end{split}
  \label{eqS:stsemi}
\end{equation}
%In our correlation measurement, signals origin from background spin decoherence, $\mathrm{^{1}H}$ in proton layer and $\mathrm{^{19}F}$ in Fomblin, their contribution to $\langle \phi ^{2} \rangle$ can be separated as
%\begin{equation}
%\langle \phi ^{2} \rangle =\langle \phi ^{2} \rangle _{\mathrm{B}} +\langle \phi ^{2} \rangle _{\mathrm{H}} +\langle \phi ^{2} \rangle _{\mathrm{F}}
%\end{equation}
%In this case, the correlation $ \langle \phi _{1} \phi _{2} \rangle $ is 
%\begin{equation}
%\langle \phi _{1} \phi _{2} \rangle =e^{-\tau _{C} \Gamma _{\mathrm{H}}} \langle \phi ^{2} \rangle _{\mathrm{H}} \ \cos( \omega _{L\mathrm{,H}}( \tau _{C} +\mathnormal{N} \tau )) +e^{-\tau _{C} \Gamma _{\mathrm{F}}} \langle \phi ^{2} \rangle _{\mathrm{F}} \ \cos( \omega _{L,\mathrm{F}}( \tau _{C} +\mathnormal{N} \tau ))
%\label{eqS:correlation}
%\end{equation}
%in which $\Gamma _{\mathrm{H}}, \Gamma _{\mathrm{F}}$ is the decoherence rate of $\mathrm{^{1}H}$ in proton layer and $\mathrm{^{19}F}$ in Fomblin. 
The parameters $ A _{\mathrm{H}} $, $ A _{\mathrm{F}} $, $T^{(^1\mathrm{H})} _{\mathrm{coor}}, T^{(^{19}\mathrm{F})} _{\mathrm{coor}}$, $\omega_{L,\mathrm{H}}  $, $\omega_{L,\mathrm{F}}  $, $\phi_\mathrm{H}$, $\phi_\mathrm{F}$ were determined by fitting the NMR signal with \Cref{eqS:stsemi}.   

\section{PL6 $T_1$ Relaxation}
\begin{figure*}
    \centering
    \includegraphics[]{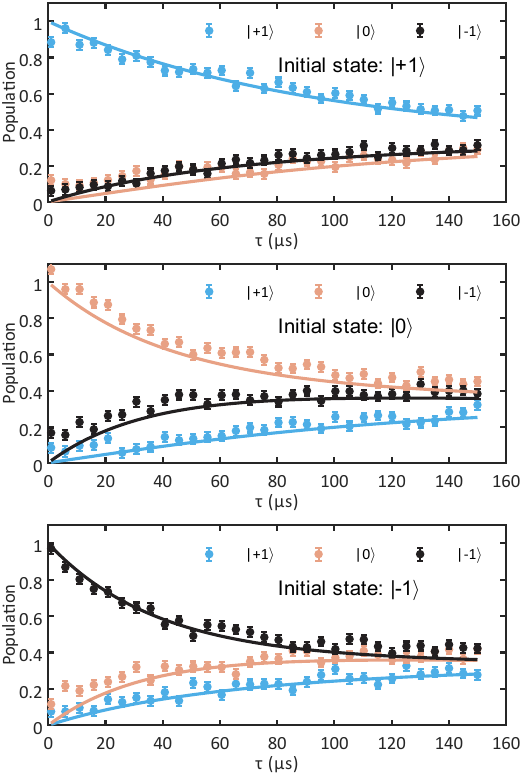}
    \caption{\textbf{Three-level longitudinal relaxation of PL6.} The relaxation behaviors of the PL6 electron spin  ($S= 1$) under 215.8 Gauss are measured with different initial states marked in each graph. The blue, brown, and  black dots stand for the states $|+1\rangle$, $|0\rangle$, and $|-1\rangle$, respectively. The solid lines are plotted by  simultaneously fitting the nine groups of data points above.}
    \label{figS:PL6_T1}
\end{figure*}
The spectral resolution in correlation spectroscopy is limited by the longitudinal relaxation time $T_1$. The dynamics of longitudinal relaxation can be described by the rate equation
\begin{equation}
\dot{n}( t) =\Gamma n( t)
\label{eqS:rate_eq}
\end{equation}
where $n(t)$ is a column vector containing the populations of $m_s=+1,0,-1$ states. The decay matrix $\Gamma$ is defined as
\begin{equation}
\Gamma =\begin{pmatrix}
-\gamma _{+1} -\gamma _{2} & \gamma _{+1} & \gamma _{2}\\
\gamma _{+1} & -\gamma _{+1} -\gamma _{-1} & \gamma _{-1}\\
\gamma _{2} & \gamma _{-1} & -\gamma _{-1} -\gamma _{2}
\end{pmatrix}
\label{eqS:Gamma}
\end{equation}
where $\gamma_{\pm1}$ is the decay rate between $m_S=\pm1$ and $m_S=0$, and $\gamma_2$ is the rate between $m_S=+1$ and $m_S=-1$. 
A set of nine experiments was conducted, covering all combinations of state preparation and readout after a variable waiting time. The results are summarized in \Cref{figS:PL6_T1}. By performing a simultaneous fit to all data points from these experiments using the methodology described in \cite{2023-9992FidelityCnotGates-Xie-Phys.Rev.Lett.}, the decay rates were determined as:
\begin{equation}
     \gamma_{+1} = 2.3(2) ~\text{kHz},\quad
         \gamma_{-1} = 11.8(5) ~\text{kHz}, \quad
          \gamma_2 = 5.0(3) ~\text{kHz}
          \label{eqS:gamma_fit}
\end{equation}
The eigenvalues $\lambda_1$, $\lambda_2$, and $\lambda_3$ of the decay matrix represent the characteristic rates of the relaxation process. Using the values from \Cref{eqS:gamma_fit}, these eigenvalues are calculated as:
\begin{equation}
\lambda_1 = 0~\text{kHz}, \quad
\lambda_2 = -10.6(8)~\text{kHz}, \quad
\lambda_3 = -27.6(12)~\text{kHz}.
\label{eqS:lambda_fit}
\end{equation}
The correlation experiment presented in Fig. 4 of the main text was performed within the subspace spanned by the $|0\rangle$ and $|-1\rangle$ states. As the population decay between these states is dominated by $\lambda_3$, the effective $T_1$ time can be estimated by
\begin{equation}
T_1 \approx {1}/{|\lambda_3|} = 36(2)~\mathrm{\mu s}.
\end{equation}
\section{PL5 as Robust Spin Sensor for NMR Spectroscopy}
\begin{figure*}
    \centering
    \includegraphics[]{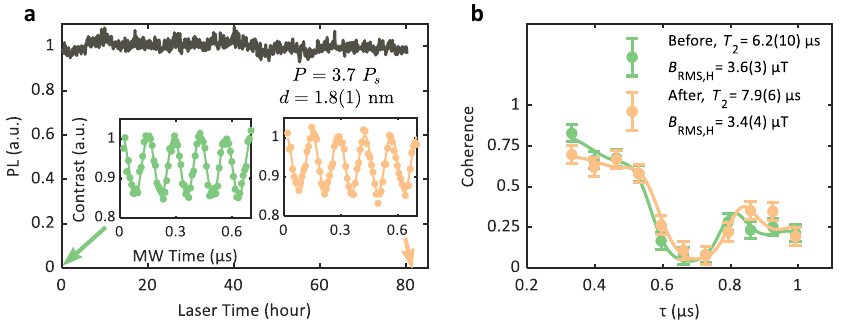}
    \caption{\textbf{PL5 performance under illumination.} \textbf{a,} Photostability of a shallow PL5 with laser power 3.7 times the saturation power $P_s$. Insets show Rabi oscillations before (green) and after (orange) illumination, demonstrating maintained optical spin readout contrast. \textbf{b,} NMR signal of $\mathrm{^{1}H}$ in immersion oil measured by the single shallow PL5 sensor in \textbf{a}. The signal is fitted with \Cref{eqS:ct1,eqS:Sg,eqS:SBNH}. Obtained $T_2, B_{\mathrm{RMS,H}}$ are displayed in right side.}
    \label{figS:Illumi_PL5}
\end{figure*}

We demonstrate that a shallow PL5 center, located at a depth of 1.8 nm, exhibits stable photoluminescence under continuous illumination for approximately 80 hours at an intensity of 3.7 times the saturation intensity (\Cref{figS:Illumi_PL5}a). The optical readout contrast remains unchanged before and after the illumination, as shown in the insets of \Cref{figS:Illumi_PL5}a. In \Cref{figS:Illumi_PL5}b, the NMR signals of $\mathrm{^{1}H}$ in immersion oil measured using this PL5 center before and after the illumination are presented. Both the coherence time and the signal amplitude show no significant variation, confirming the exceptional photostability and consistent performance of the sensor throughout the prolonged experiment.

\bibliography{main_bibliography}% common bib file
%% if required, the content of .bbl file can be included here once bbl is generated
%%\input sn-article.bbl

%% Default %%
%%\input sn-sample-bib.tex%